\documentclass{iopart}
\usepackage{graphicx}
\begin{document}
\title[Color Superconductivity and Radius of Quark Star in Extended NJL
Model]
{Color Superconductivity and Radius of Quark Star
 in Extended NJL Model by Using the Dimensional Regularization.}
\author{
T. Fujihara$^1$,T. Inagaki$^2$, D. Kimura$^1$
}
\address{$^1$
Department of Physics, Hiroshima University,
Higashi-Hiroshima, Hiroshima
739-8526, JAPAN}
\address{$^2$Information Media Center,Hiroshima University, 1-7-1
 Kagamiyama, Higashi-Hiroshima 739-8521,Japan}
\begin{abstract}
A radius of a dense star on the color superconducting phase is investigated
in an extended NJL type model with two flavors of quarks.
Since the model is non-renormalizable, the results depend on the 
regularization procedure. Here we apply the dimensional 
regularization and evaluate the radius of a dense star.
Evaluating the TOV equation, we show the relationship 
between mass and radius of the dense star in the dimensional regularization.
\end{abstract}
\pacs{12.38.Aw,25.75.Nq,26.60+
%\hspace{30mm}HUPD0509
}

\section{Introduction}
Quarks and gluons matter is expected to have a 
new phase at high density.
The phase is called ``Color Superconductivity (CS)`` phase,
see Refs.\cite{Bailin:1983bm}-\cite{Alford:1998mk}.
In the CS phase diquark pair  construct a Cooper pair like state,
therefore the color SU(3) symmetry is broken down.
Since the CS phase appears at extremely high density,
it is difficult to observe the phase in a Laboratory.
However, there is a possibility to find the phase in
astrophysical observations.
In the core of dense stars, like neutron stars, quark stars and so on.
The CS phase may be realized.
Thus much attention has been paid to such dense stars
to find a evidence of the color superconductivity's.
The observable quantities of such dense stars are mass and radius.

A constraint for the mass and radius of the star is found  by solving the
balance equation between the force of gravity and the pressure of the matter 
inside.
The balance equation is represented by
the Tolman-Ophehimer-Volkov (TOV) equation. 
The solution is determined by the equation of state (EoS)
of the quark matter inside the star. 
It was pointed out that the existence of the CS phase
may decrease the minimum of radius of such dense stars
\cite{Alford:1997zt}-\cite{Shovkovy:2003ps}.

Of course the EoS depends on the model
to describe the quark matter.
One of the often used model is Nambu-Jona-Lasinio (NJL) model
in which the chiral symmetry is broken down dynamically \cite{NJL}.
The model reproduces a lot of phenomena in the hadronic phase of QCD.
We use the extended NJL model to include the interactions between diquark and
analyzed it at finite chemical potential.

Since this model is non-renormalizable,
we must regularize the model to obtain the finite result.
However, most of analysis has been done with a cut-off regularization.
It is achieved by introducing the cut-off scale as a upper bound for the
momentum integral.
The scale can be determined phenomenologically. 
However, the cut-off scale often breaks the symmetry of the model.
Furthermore the critical chemical potential 
where the color superconductivity takes place is the same order 
with the cut-off scale. In such a situation it is expected that the 
regularization procedure has non-negligible effects on
the precise analysis of the physics in CS phase.
Therefore we launched our plan to analyze the extended NJL model by using 
an other regularization.
There are a lot of procedure to regularize the model.
The usual NJL model has been studied by the dimensional regularization
under some extreme conditions, especially in curved spacetime
\cite{Inagaki:1997kz}.
In the dimensional 
regularization we analytically continue the model to the spacetime dimensions 
less than four. Then we can keep most of the symmetries and regularize the 
model.

In the present paper we study the phase 
structure of the extended NJL model in the dimensional regularization.
We calculate the effective potential in dimensions less than four.
The spacetime dimensions corresponds to a regularization
parameter which should be determined phenomenologically.
We solve the gap equation TOV equation simultaneously,
and show the relationship between radius and mass of the dense star.

\section{Extended NJL model}
We consider the low energy effective theory of QCD with two flavors of 
quark fields.
To study the CS phase the NJL model is extended to include
the diquark interactions.
The extended NJL model is defined by the Lagrangian density \cite{EKK}
\begin{eqnarray}
 \mathcal{L} &=& \bar{\psi}_{ai}i \partial\!\!\!/ \psi_{ai}
    + G_S\{(\bar{\psi}_{a j}\psi_{a j})^2
    + (\bar{\psi}_{aj}i\gamma_5 {\tau}_{jk} \psi_{ak})^2\} \nonumber \\
   &&
    + G_D(i\bar{\psi}^c_{aj}\varepsilon_{jk}
    \epsilon^b_{ad} \gamma_5\psi_{d k})
   (i\bar{\psi}_{f l}\varepsilon_{lm}\epsilon^b_{f g} \gamma_5\psi^c_{g m}) ,
\label{lag}
\end{eqnarray}
where the indexes $a,b,d,e,f,g$ and the indexes 
$j,k,l,m$ denote the colors and flavors of the fermion fields, 
${\tau}_{jk}$ represents the isospin Pauli matrices, $\psi^c$ 
is a charge conjugate of the field $\psi$ and $G_S$ 
and $G_D$ are the effective coupling constants. 
The second line in Eq.(\ref{lag}) corresponds to the diquark 
interactions.
Introducing the auxiliary fields, chemical potential $\mu$ and
temperature $\beta\equiv 1/T$,
the Lagrangian density is simplifies to
\begin{equation}
 \mathcal{L}_{aux} =
  \frac{1}{2}\bar{\Psi}G^{-1}\Psi 
  -\frac{1}{4G_S}(\sigma^2+\pi^2)
  -\frac{1}{4G_D}|\Delta^b|^2,
\label{eq:Laux_NGform}
\end{equation}
where $G$ represents the quark propagator in Nambu-Gor'kov representation
\begin{equation}
 G^{-1} \equiv
  \left(
   \begin{array}{c}
    i\partial\!\!\!/-i\mu\gamma_4-\sigma -i\gamma_5
     \pi\cdot\tau \qquad
     -i\Delta^a\varepsilon\epsilon^a\gamma_5 \\
    -i{\Delta^b}^*\varepsilon\epsilon^b\gamma_5 \qquad
     i\partial\!\!\!/+i\mu\gamma_4-\sigma - i\gamma_5
     {\pi}\cdot{\tau}^T
   \end{array}
  \right),
\end{equation}
$\Psi$ is an eight components spinor defined by
\begin{equation}
\Psi=\left(\begin{array}{c}\psi \\ \psi^c\end{array}\right).
\end{equation}
From the equation of motion for the auxiliary fields we obtain the
following correspondences,
$\sigma\sim -G_s \bar{\psi}_{aj}\psi_{aj}$, 
$\pi\sim -G_s \bar{\psi}_{aj}i\gamma_5{{\tau}}_{jk}\psi_{ak}$,
$\Delta^b\sim -G_D i\bar{\psi}^c_{aj}\varepsilon_{jk}\epsilon^b_{ad}
\gamma_5\psi_{d k}$.
We assume $\pi=0,\Delta^{1,2}=0,\Delta\equiv\Delta^3$ in what follow,
because the model has chiral symmetry and
we can chose the direction of color symmetry breaking arbitrarily.

To define a finite theory we adopt renormalization conditions,
\begin{eqnarray}
 \frac{1}{2G_S^r}=
   \frac{\partial^2 V_{eff}}{\partial\sigma\partial\sigma}~,~
 \frac{1}{2G_D^r}=
  \frac{\partial^2 V_{eff}}{\partial\Delta\partial\Delta}
  \label{ex:renormal_condition},
\end{eqnarray}
where right-hand side is evaluated at $T=0$,$\mu=0$.
We assumed $\sigma=\sigma_0$ $\Delta=0$ at $T=0$,$\mu=0$.
We assumed the relation between $G_S^r$ and $G_D^r$
as $G_D^r=3/4 G_S^r$.
We set the renormalization scale $M$ equal to $\sigma_0$.
The coupling constant $G_S$ is fixed
to produce consistent values for
the pion mass $m_\pi$ and the decay constant $f_\pi$.
Then parameter $G_s$ is written as a function
of dimension $D$.

To see the phase structure of the extended NJL model at finite temperature
and chemical potential we evaluate the effective potential
in the mean-field approximation.
We introduce the temperature and the chemical potential to the theory by
the imaginary time formalism.
Performing the integration over the fermion fields $\psi$ and $\bar\psi$
and replacing the auxiliary fields with their expectation value in 
Eq.(\ref{ex:pot}), we obtain the effective potential, $V_{eff}$, of the model,
\begin{eqnarray}
V_{eff}(\sigma, \Delta)
 &=&\frac{\sigma^2}{4G_S}+\frac{|\Delta|^2}{4G_D}
  - \frac{1}{ 2 \beta \Omega }\ln\det\left(G^{-1}\right)\!\label{ex:pot}.
\end{eqnarray}
where $\Omega$ denotes the volume of the system, 
$\Omega \equiv \int d^{D-1} x$.

The auxiliary field $\Delta$ is order parameter of
color SU(3) symmetry.
Therefore non-zero expected value of $\Delta$ mean
color SU(3) symmetry breaking and realized the SC phase.
The expectation values for $\sigma$ and $\Delta$ are found by observing
the minimum of this effective potential. To find the minimum we numerically 
calculate the effective potential (\ref{ex:pot}) in D-dimensions. The 
expectation values are shown as a function of the chemical potential in 
Fig.1.
We also evaluate the effective potential and calculate
the expectation values in a naive cut-off regularization.
The behavior of the expectation value for $\sigma$
with the dimensional regularization
as well as with the cut-off regularization.
However, the expectation value for $\Delta$
shows significantly different behavior for large chemical potential.
As is clearly seen in Fig.1, the expectation value for $\Delta$
is a monotone increasing function of the chemical potential
in the dimensional regularization.
In the cut-off regularization $\Delta$ is suppressed as the chemical
potential approaches the cut-off scale.

The energy-density and the pressure of the system is given by the
energy-momentum tensor.
In the case of spherical and static space-time
relationship between the energy-momentum tensor,
energy-density $\rho$ and the pressure $P$
 is given by
$ T_{\mu\nu} = diag(-\rho,P,P,P)$.
Thus we calculate the energy-momentum tensor at finite $T$ and $\mu$
in the dimensional regularization.

\section{Radius and mass of dense stars}
To see the physical implication of the regularization dependence
we discuss the radius and the mass of dense stars.
Inside the star the gravitational force should be balanced with the 
pressure of the matter. For spherically symmetric stars it is expressed 
by TOV equations \cite{Oppenheimer:1939ne}.
Thus we assume that the CS phase is realized inside
the stars and solve the TOV equations defined by 
\begin{eqnarray}
\hspace{-3em}
 \frac{d P(r)}{d r}
  =
  -GM(r)
  \frac{\left( \rho(r) + P(r)\right)
  \left( 1+ \frac{4\pi r^3P(r)}{M(r)}\right)
  }{r\left(r -2GM(r)\right)},~
 \frac{d M(r)}{dr}
  &=& 4\pi r^2 \rho(r),
\end{eqnarray}
where pressure $P(r)$ ,energy density $\rho(r)$ and mass of star $M(r)$ 
as a function of the radial distance $r$ from the center of a star.
In Fig.2 we draw the relation between radius and mass of the star.
The result at $D=2.5$ shows a similar behavior to the one
obtained by cut-off regularization.
Since the convergence of the numerical calculation becomes slower
for smaller $M$ at $D=2.5$,
we can draw the solution for larger $M$.
\begin{figure}[t]
\begin{center}
\begin{minipage}{0.45\textwidth}
\includegraphics[height={!},width={\textwidth}]{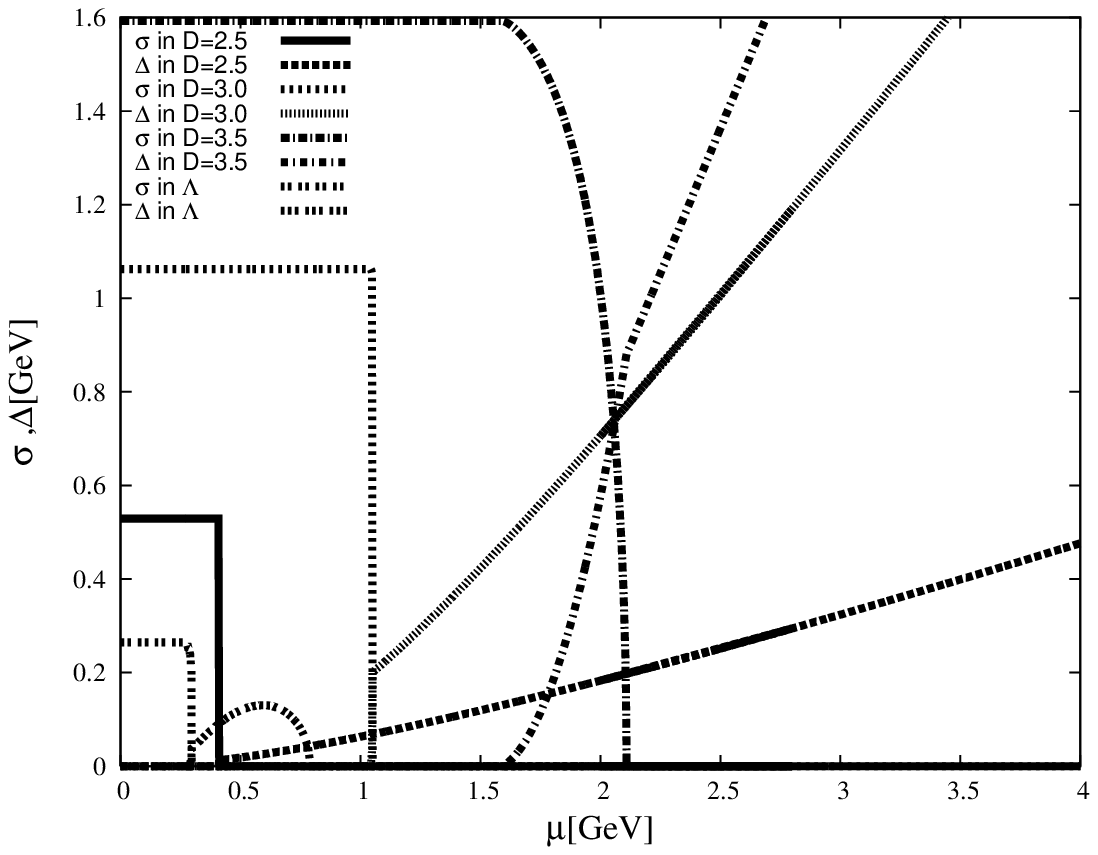}
\caption{Behaviors of the expectation
 value of auxiliary fields $\sigma$ and $\Delta$.
The cutoff is equal to $0.73$GeV.}
\end{minipage}
\begin{minipage}{0.45\textwidth}
\includegraphics[height={!},width={\textwidth}]{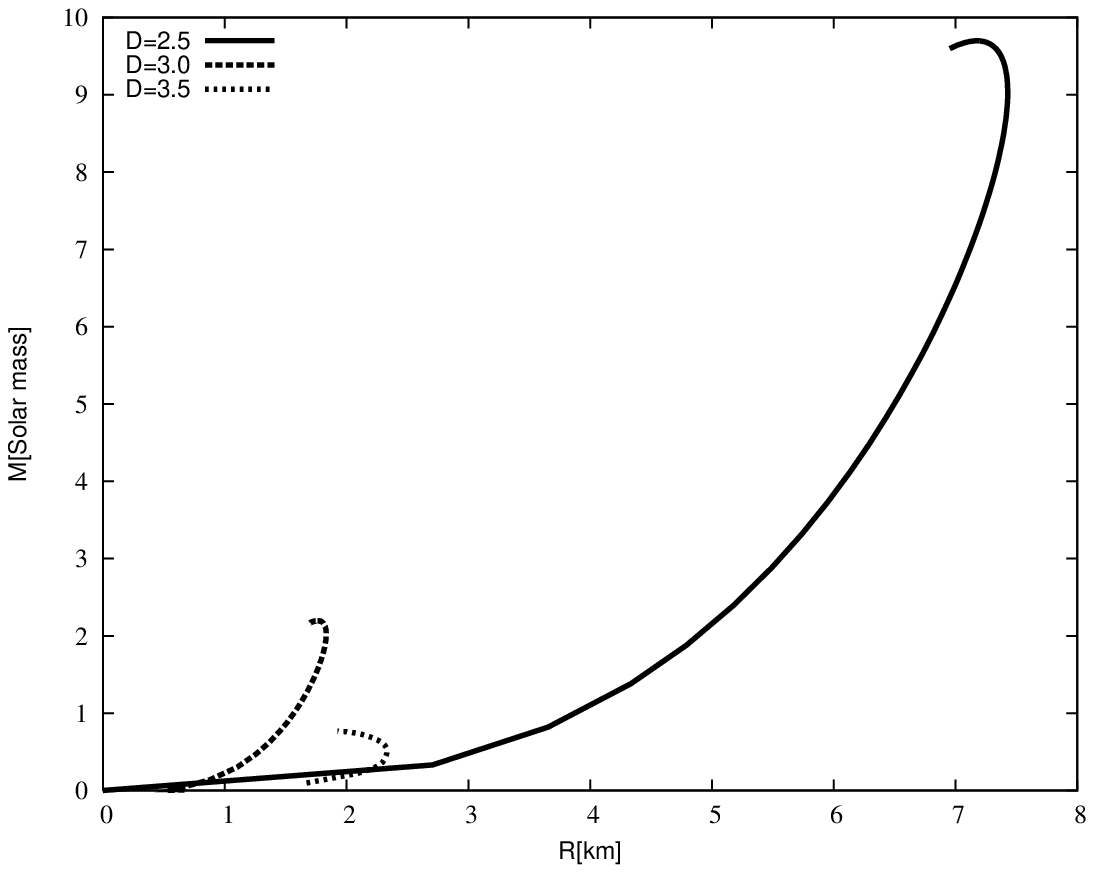}\\
\caption{Relationship between mass and radius with dimensional
 regularization.}
\end{minipage}
\end{center}
\end{figure}

\section{Conclusion}
We have investigated the extended NJL model in dimensional regularization.
Evaluating the effective potential we show the behavior of the
mass gap $\sigma$ and $\Delta$ as a function of the chemical potential.
The mass scale for $\sigma$ and $\Delta$ significantly depends on
the spacetime dimensions.
EoS of the model is found from the expectation value of the stress tensor.
By using the EoS we numerically solve the TOV equation and show the
relationship between radius and mass of the dense star.
It also has significant dependence on the dimension.
It is expected that the significant dependence disappears
under a suitable renormalization condition.

There are some remaining problems.
To make a more realistic model we should consider
the effect of the strange quark.
We  should also impose the conditions to satisfy the
color neutrality and electromagnetic neutrality.
It is not imposed in the present paper.
As is known,
the condition may affect the ground state and the physical
behavior of the dense star\cite{Shovkovy:2003uu}.


\section*{References}
\begin{thebibliography}{99}
\bibitem{Bailin:1983bm}
  Bailin D and Love A 1984 {\it Phys. Rept.} {\bf 107} 325
\bibitem{Alford:1997zt}
  Alford M, Rajagopal K and Wilczek F 1998 \PL B {\bf 422}
  247
\bibitem{Rapp:1997zu}
  Rapp R, Schafer T, Shuryak E V and Velkovsky M 1998
  \PRL {\bf 81}  53
\bibitem{Alford:1998mk}
  Alford M, Rajagopal K, Wilczek F 1999 \NP {\bf B537} 443
\bibitem{Shovkovy:2003ps}
  Shovkovy I, Hanauske M and Huang M 2003 \PR D{\bf 67} 103004
\nonum  Shovkovy I, Hanauske M and  Huang M 2003 {\it eConf} C{\bf 030614} 039
 ({\it Preprint} hep-ph/0310286)
\bibitem{NJL}
  Nambu Y and Jona-Lasinio G 1960 \PR {\bf 122} 345
\nonum  Nambu Y and Jona-Lasinio G 1961 \PR {\bf 124} 246
\bibitem{Inagaki:1997kz}
Inagaki T, Muta T and Odintsov S D 1997 
Prog.\ Theor.\ Phys.\ Suppl.\  {\bf 127} 93 
%[arXiv:hep-th/9711084].
\bibitem{EKK}
  Ebert D, Kaschluhn L and Kastelewicz G 1991 \PL B {\bf 264} 420
\bibitem{Oppenheimer:1939ne}
  Oppenheimer J R and Volkoff G M 1939 \PR {\bf 55} 374
\bibitem{Shovkovy:2003uu}
Shovkovy I and Huang M 2003
\PL B {\bf 564} 205
%[arXiv:hep-ph/0302142].
\end{thebibliography}
\end{document}